# Bond Stiffness, not Chain Length, Dictates Polymer Infiltration into Nanopores

David J. Ring, Robert A. Riggleman* and Daeyeon Lee*

*Chemical and Biomolecular Engineering, University of Pennsylvania, Philadelphia, PA 19104*

We study the effect of physical confinement on the capillary infiltration of polymers into cylindrical nanopores using molecular dynamics simulations. In particular, we probe whether the critical contact angle above which capillary rise infiltration ceases to occur changes for long chain polymers, possibly due to loss of conformation entropy induced by chain confinement. Surprisingly, the critical contact angle does not strongly depend on the length of polymer chains and stays constant for large N. A free energy model is developed to show that the critical angle for infiltration depends strongly on the size of statistical segments rather than the total chain length, which we confirm by performing MD simulations of infiltration with semi-flexible polymers. These results could provide guidelines in manufacturing polymer nanostructures and nanocomposites using capillary rise infiltration.

Capillarity-driven flow of polymers in cylindrical tubes with nanoscale pores enables fabrication of novel nanostructures such as nanotubes and nanorods and at the same time provides a versatile method of confining polymer chains under physical confinement [1,2]. Prior work has reported changes in the translational and segmental dynamics of polymers in cylindrical nanopores [3] as well as changes in glass transition temperature of semicrystalline polymers [4]. Cylindrical confinement also has a significant impact on the phase behavior of block copolymers [5] and semicrystalline polymers [6]. Many of these observed changes in the morphology and dynamic behavior of polymers under nanoscale confinement is attributed to the changes in polymer configurations [7,8]. That is, when the characteristic dimension of unperturbed polymer chains such as their end-to-end distance ($R_e$) is comparable to or greater than the diameter of the pores, chains lose significant entropy, affecting their behavior [9].

Experimental results as well as molecular dynamics (MD) simulation of capillary imbibition of polymers into nanoscale pores have shown that infiltration dynamics can be described using the Lucas-Washburn theory, $h = \sqrt{\frac{D\gamma \cos\theta}{4\eta} t}$, which describes the capillary infiltration dynamics of liquids with viscosity $\eta$ and surface tension $\gamma$ wicking into pores of diameter D [10,11]. The critical parameter that determines whether a liquid would undergo capillarity-driven imbibition into a pore is the contact angle $\theta$ of the liquid on the pore surface. The model indicates that as long as $\theta$ is less than 90°, the liquid will undergo capillary-driven flow into a pore. For polymers where the pore diameter is smaller than $R_e$, there is an intriguing possibility that the critical angle for infiltration to occur could be less than 90° since the polymer chains would require additional thermodynamic driving force (i.e., more favorable wetting) to overcome loss of conformational entropy.

In this study, we use molecular dynamics (MD) simulations to investigate how physical confinement of polymers in nanoscale cylindrical pores impacts the capillarity-driven infiltration of polymers into the nanopores. In particular, we study the impact of chain confinement on the critical contact angle ($\theta_c$) above which infiltration of polymers into nanopores ceases to occur. To enable this study, we find the relationship between the contact angle of polymer on the solid surface and polymer-surface interactions for a range of polymer chain lengths. By varying the pore diameters, polymer chain length and statistical segment sizes, we systematically probe the effect of confinement on infiltration of polymers into cylindrical nanopores. Somewhat unexpectedly, the critical contact angle does not strongly depend on the confinement ratio $(R_e/D)^2$ as the size of the chains becomes comparable to or larger than the pore size. A simple free energy argument shows that the critical angle for infiltration depends strongly on the size of statistical segments rather than the total chain length, consistent with our simulation results.

We use a bead-spring model with Lennard-Jones particles, harmonic bonds, and harmonic bending potentials to model our coarse-grained polymers [12–14]. By adding the angle potential $u(\theta) = \frac{k_\theta}{2}(\theta - \theta_0)^2$ we can model both fully flexible chains ($k_\theta = 0$) and semi-flexible chains ($k_\theta = 10$) having a bond angle of $\theta_0 = 2\pi/3$. This allows us to simulate polymers with different statistical segment sizes to investigate the effect of backbone rigidity on the critical contact angle of capillary rise. All simulations are performed at $k_B T/\varepsilon = 0.7$ unless otherwise specified.

All polymer melts are formed from collections of overlapping chains with approximately Gaussian statistics [12]. Overlaps are removed by ramping a soft cosine interaction potential in place of the standard LJ potential before equilibrating the polymer melts with their actual interaction potentials [15]. All melts are equilibrated such that they diffuse on average by their chain dimension, $R_e$, which can be calculated from a comparatively short simulation. All surfaces are composed of LJ sites arranged in a triangular lattice with lattice parameters l=(0.64,1.11). Non-bonded interactions are controlled with the cut and shifted LJ potential $u_{ij}(r) = 4\varepsilon_{ij}\left[\left(\frac{\sigma_{ij}}{r}\right)^{12} - \left(\frac{\sigma_{ij}}{r}\right)^6\right] - 4\varepsilon_{ij}\left[\left(\frac{\sigma_{ij}}{r_c}\right)^{12} - \left(\frac{\sigma_{ij}}{r_c}\right)^6\right]$ where $r_c = 1.75$ is the cut-off distance and $\varepsilon_{ij}$ represents the interaction strength between atoms i and j.

Prior to performing the infiltration simulations, we reduce $\varepsilon_{PC}$ between the LJ particles that comprise the polymer (P) monomers and the capillary (C) sites to a non-wetting value so that the polymer-vapor interface can relax near the opening of the capillary. Subsequently, we increase $\varepsilon_{PC}$ and test whether we observe spontaneous infiltration as shown in Figure 1a. The height of the melt in the capillary is calculated by fitting $\rho(z) = \frac{\rho_l+\rho_v}{2} - \frac{\rho_l-\rho_v}{2}\tanh\left[\frac{2(z-z_0)}{l}\right]$ to the density profile, as shown in Figure 1b, where $\rho_l$ is the density of the polymer melt in the capillary, $z_0$ is the height of the interface, and $l$ is the thickness of the interface. The polymer-vapor interface has a thickness of $1-2\sigma$. The polymer height (h) is roughly proportional to time (t) as shown in Figure S2 in the supplemental information, which is consistent with the inertial regime of capillary imbibition [16]. Our focus is to identify the critical angle above which infiltration of polymers into nanopores ceases to occur. Thus, we run each simulation for at least a diffusion time, $t/\tau_{LJ} = R_e^2/\mathcal{D}$ where $\mathcal{D}$ is the bulk diffusivity of a polymer of length N and $\tau_{LJ} = \sigma\sqrt{m/\varepsilon}$ where $\tau_{LJ}$ is the LJ timescale calculated from LJ parameters $\varepsilon$ and $\sigma$, and the mass, $m$. On these timescales, the polymers in the film below the capillary are able to relax sufficiently that we expect our simulations to not be limited by kinetic traps. As discussed below, results using this protocol agree with calculations of the free energy change during infiltration.

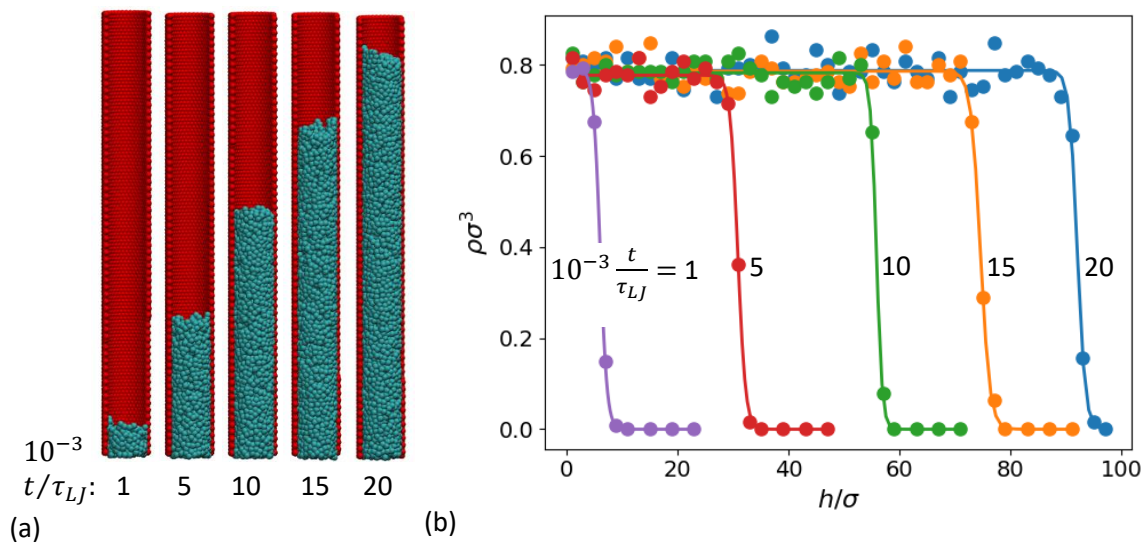

FIG. 1. (a) A cylindrical capillary with diameter D=9, height H=100, and polymer chains of length N=10 are used to simulate the process of polymer wicking. At $10^{-3} t/\tau_{LJ} = 1$, the polymers are held outside the capillary by unfavorable

interactions but can still fluctuate as a free surface at the mouth of the capillary. For t > 0, the interactions between capillary and polymers are turned up to induce wicking into the capillary. (b) the height of the polymer interface can be tracked by fitting a sigmoidal function to $\rho(z)$. The full system showing the polymer film beneath the pore is shown in Supplemental Information.

The strength of interactions between a solid surface and a liquid at the macroscopic scale is often expressed in terms of contact angle, $\theta$, which in our model is primarily controlled through the LJ parameter $\varepsilon_{PC}$. To correlate these two parameters, we determine the contact angle of a polymer droplet with a pseudo-infinite cylindrical geometry [17] on the solid surface with the same lattice structure as the cylindrical pores as shown in the inset of Figure 2a. The use of cylindrical geometry removes the curvature of the contact line so that the contact angle determination is not affected by line tension and in turn remains independent of the size of the sessile drop. All simulations start with a melt droplet contact angle close to 90° and are annealed until the contact angle reaches a constant value.

We observe two prominent trends in the contact angle shown in Figure 2a. First, $\cos\theta$ is proportional to $\varepsilon_{PC}$ for all chain lengths, N, which is expected as stronger enthalpic interactions between polymer chains and the solid surface will induce stronger wetting (i.e., smaller $\theta$). Interestingly, for a given value of $\varepsilon_{PC}$, $\theta$ increases ($\cos\theta$ decreases) for increasing N.

As the physical reason behind the observed dependence of polymer contact angle on N for a given $\varepsilon_{PC}$ is not obvious, we further investigate the role of the interfacial tension of the free polymer surface in dictating the contact angle. We calculate the surface tension from the asymmetry of the pressure tensor in a free-standing, planar thin film simulation [14,18] as explained in the supplemental information. It is known from previous work on the concentration of chain ends at free interfaces that the surface tension can depend on chain length and saturates for long chains [19,20]. It was previously reported that the segregation of the chain ends to interfaces leads to this entropic contribution to the surface tension, and the dependence of the interfacial tension on N has a form that does not depend on the details of the molecular model [21].

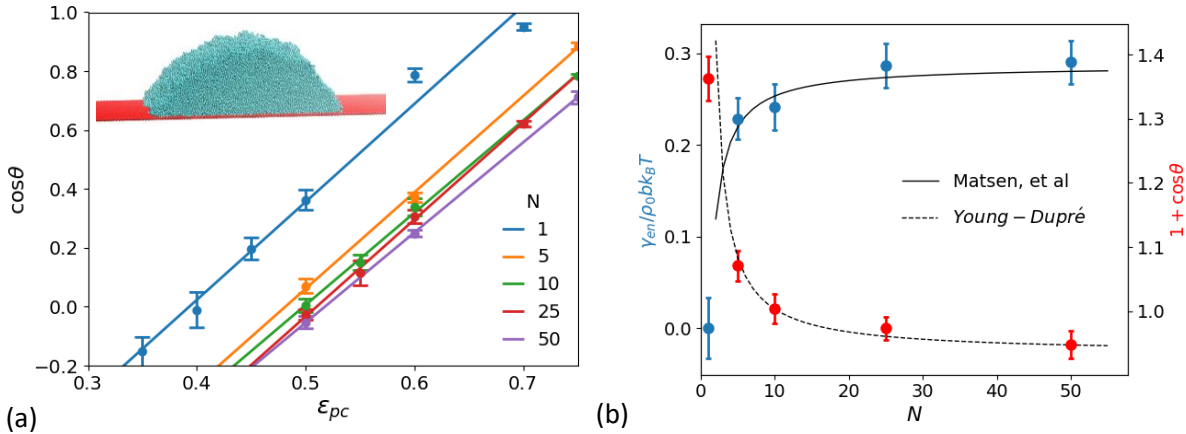

FIG. 2. (a) Contact angle simulations are used to parameterize the relationship between the interaction energy ($\epsilon$) and contact angle ($\theta$). $\epsilon$ varies approximately linearly with $\cos\theta$ for all values of $N$. (b) The dependence on chain length $N$ is explained by the molecular weight dependence of the surface tension ($\gamma$) from Ref [21] and the work of adhesion from the Young-Dupré equation. The work of adhesion measured is $W_{ad} = 0.155\ \epsilon/\sigma^2$. Literature [22] reports similar values for LJ polymers.

This entropic component of surface tension, $\gamma_{en}$, when added to the simple fluid surface tension, $\gamma_1$, yields the total surface tension, $\gamma$. Thus, we plot surface tensions in Figure 2 from simulation by first computing $\gamma - \gamma_1$ with the simulated values, where $\gamma_1$ was calculated for a film of a simple LJ fluid. We expect to this quantity be comparable to the theoretical $\gamma_{en}$ presented in Ref. [20] and is given as

$$\frac{\gamma - \gamma_1}{b\rho_0 k_B T} \approx \frac{\gamma_{en}}{b\rho_0 k_B T} = \Gamma_\infty - \frac{2A}{N}, \quad (1)$$

where $\Gamma_\infty$ and $A$ are constants previously computed [21] for polymers with Hookean spring bonds; they can be interpreted as the surface tension of infinitely long polymers and the contribution from a single chain end, respectively. The other constants $b$, $\rho_0$, and $k_B T$ are the Kuhn length, monomer density, and temperature, respectively, which are taken from our system. We observe in Figure 2b that the entropic component of our surface tension is in good agreement with Eq 1, even though we use the parameters $\Gamma_\infty$ and $A$ taken from the calculations in Ref [21].

Rearranging and inserting Eq 1 into the Young-Dupré equation provides an expression for the N dependence of the contact angle,

$$1 + \cos\theta \approx \frac{W_{ad}}{\gamma_1 + \gamma_{en}} = \frac{W_{ad}}{\gamma_1 + b\rho_0 k_B T \left(\Gamma_\infty - \frac{2A}{N}\right)} \quad (2)$$

where $W_{ad}$ is the work of adhesion between the polymer and surface [23]. Fitting equation 2 to the contact angle data in Figure 2 by varying $W_{ad}$ at $\varepsilon_{PC} = 0.5$ gives $W_{ad} = 0.155$, which is in good agreement with previous simulation studies [22] of similar polymer models. This appears to imply that chain ends influence the contact angle and that the saturation of the contact angle results from a reduction in the number of chains at the solid-polymer interface and free surface. However, the work of adhesion remains nearly constant. This appears to be due to a balance between $\gamma_P$ and $\gamma_{PC}$ which offsets the effect of chain ends.

We probe the effect of confinement on infiltration of polymer chains into nanopores by varying both the chain length and capillary diameter and determining the critical contact angle for infiltration to occur by varying the interactions of the polymer with the capillary walls, $\varepsilon_{PC}$. The extent of confinement is quantified with a confinement ratio $\delta = \frac{R_e^2}{D^2}$ where D is capillary diameter. To distinguish whether a particular set of conditions leads to spontaneous infiltration of the polymers, we expect that the melt should infiltrate to a height larger than the polymer's end-to-end distance within a diffusion time, $\tau_e \sim R_e^2 / D$. When the height inside the capillary increases at such a rate, we interpret the results as implying that any observed infiltration is driven by capillarity and not governed by a diffusion process. For example, in Figure 3 we plot the infiltrated height of N = 50 polymer infiltrating a pore of D = 5 as a function of time; we observe that only for a contact angle $\theta \leq 82°$, infiltration is observed.

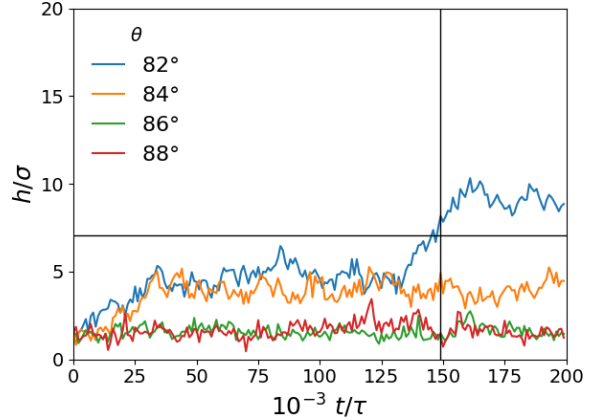

FIG. 3. Height trajectories of fully-flexible polymer chains of length $N = 50$ show that the polymers only infiltrate past their mean squared end-to-end distance (black line) if $\theta < 84°$. The pore diameter D is $5\sigma$.

In Figure 4a, we show the calculated critical contact angles by identifying chains that are undergoing infiltration under varying pore and chain sizes. For flexible chains, we observe a suppression of the critical contact angle from the simple fluid to small N. However, $\theta_c$ reaches a plateau at approximately 83° and does not show an observable dependence on $N$ for $N \geq 25$. Given that the entropy of an ideal chain scales with chain length, $\Delta S \sim R_e^2 \sim N$, and that confinement causes a loss of entropy proportional to $N$ [7], it is initially surprising that $\theta_c$ does not depend on N for large N or strong confinement as shown in Figure 4b. To verify our empirical definition of $\theta_c$, we have calculated the free energy of infiltration using umbrella sampling for select values of $N$. The umbrellas are implemented by biasing the separation between centers of mass between the surface containing the capillary and the polymer melt with a harmonic potential. The critical contact angle determined based on free energy calculations for two values of $N$ (Figures 4a and 4b) are slightly larger than those obtained based on the diffusion length criteria, and more importantly the values of $\theta_c$ show little dependence on $N$ for $N > 25$. The small difference in $\theta_c$ obtained using the two approaches we believe is likely due to a kinetic effect near the entrance of the capillary. More detailed discussion on umbrella sampling is provided in the supplemental information.

To understand this trend, we develop a free energy model accounting for contributions from surface free energy and the confinement free energy [7] similar to the model presented recently [9]. As shown schematically in Figure 4c, the change in surface area with a change in height $dh$ is $dA = \pi D dh$ and thus the gain in surface energy can be expressed as $(\gamma_C - \gamma_{PC})dA$ where $\gamma_C$ is the capillary surface energy and $\gamma_{PC}$ is the polymer-capillary interfacial energy. The differential change in the confinement free energy as the fluid height increases can be expressed by taking into account the loss of conformational entropy $k_B T \left(\frac{Nb^2}{D^2}\right)\rho dV$, where $\rho = \rho_0/N$ is the density of polymer chains, and $dV = \frac{\pi D^2}{4} dh$ is the increase in volume of the polymer in the cylindrical tube. Combining the entropic penalty with the contributions from the interfacial tension, the total change in the free energy upon a change in height $dh$ is the sum of these two contributions, which gives

$$dF = (\gamma_C - \gamma_{PC})\pi D dh + kT\left(\frac{Nb^2}{D^2}\right)\rho \frac{\pi D^2}{4} dh. \quad (3)$$

By substituting in Young's equation [23], $\cos\theta = (\gamma_{PC} - \gamma_C)/\gamma_P$, converting the chain density, $\rho$, to the monomer density, $\rho_0 = \rho N$, and setting the free energy change $dF/dh = 0$, an analytical solution for the critical contact angle can be obtained as

$$\cos\theta_c = \frac{\rho_0 k T b^2}{4\gamma_P D} \quad (4)$$

Remarkably, Eq. 4 suggests that the critical contact angle does not depend on N, consistent with our simulations results for $N > 25$. This result can be rationalized by considering the competing effects of chain entropy and the number of chains in the cylindrical pore. Although increasing the confinement of a single chain results in a loss of entropy, this change is compensated by the number of chains confined in the cylindrical pore such that the total melt entropy remains constant. Equation 4 also predicts that critical contact angle should depend on Kuhn length, $b$, such that a polymer with a larger Kuhn length would result in smaller $\theta_c$.

To test this prediction, we perform infiltration simulations with a semi-flexible model polymer that has a larger Kuhn length, which are produced by adding the angle potential $u(\theta) = \frac{k}{2}\left(\theta - \frac{2\pi}{3}\right)^2$ along the backbone of the polymer chain [13]. Calculating the persistence length from the decay of the bond-autocorrelation function (see supplemental info), we show that these semi-flexible chains with $b \approx 2l_p \approx 2.62\sigma$ are significantly stiffer than the flexible chains where $b \approx 1.24\sigma$. The critical contact angles calculated for these semi-flexible chains using the kinetic and the umbrella sampling approaches are also shown in Figure 4a. As was the case with the flexible chain, the critical contact angle shows a reduction up to $N \approx 25$ and then plateaus around $77°$, indeed lower than that of the flexible chains.

We can analytically predict the value of the plateau in $\cos\theta_c$ by substituting equation 1 into equation 4 and taking the limit for large N to find

$$\cos\theta_c = \frac{b/D}{4\left(\frac{\gamma_1}{\rho_0 b k_B T} + \Gamma_\infty \left(1 - \frac{2A}{\Gamma_\infty N}\right)\right)} \quad (5)$$
$$\approx \frac{b/D}{4(\Gamma_1 + \Gamma_\infty)}$$

where $\Gamma_1 = \frac{\gamma_1}{\rho_0 b k_B T}$. We observe from Eq 5 that dimensionless surface tensions for both simple fluid ($\Gamma_1$) and infinite chains ($\Gamma_\infty$) and the ratio $b/D$ are sufficient to characterize $\theta_c$ in the long-chain limit. Since in our case the monomeric fluid has the same dimensionless surface tension $\Gamma_1$ for both polymer models and if we assume $\Gamma_\infty$ is insensitive to small changes in the backbone stiffness, then the changes in $b$ accounts for the plateaus of $83°$ and $77°$ observed in $\theta_c$ in Figures 4a and 4b. Thus, the more rigid polymer backbones decrease the critical contact angle for infiltration more than a flexible polymer backbone.

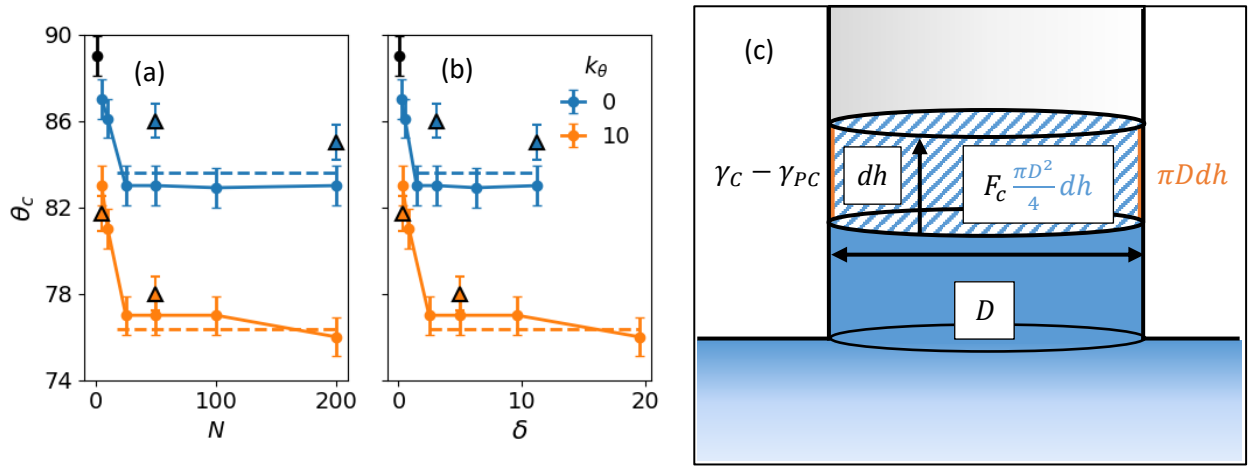

FIG. 4. The critical angle for infiltration of flexible chains plateaus around a value of 83° whereas the critical value for stiff chains plateaus around 77°. The dashed lines indicate theoretical predictions from the energy balance.

In this work, we have shown that despite the confinement-induced loss of entropy that is experienced by individual polymer chains, the critical contact angle that induces spontaneous surface-driven infiltration of polymers into cylindrical pores do not strongly depend on N, for large N. The polymer melt instead tends to compensate for the entropy loss of individual polymers in such a way to keep the total change in free energy constant. The stiffness of the chain, as represented in term of the Kuhn segment length, rather than the overall length of the chain determines the critical contact angle. The implications of these findings are that stiffer chains have a smaller window of contact angles under which they can infiltrate. Additionally, studying model cylinder geometries provides insight for polymer imbibition into other porous media such as random packings of nanoparticles, which have paved the way towards new fabrication methods of composite materials with extremely high volume fractions of nanoparticles [24].

**References:**

[1]  M. Zhang, P. Dobriyal, J. T. Chen, T. P. Russell, J. Olmo, and A. Merry, Nano Lett. **6**, 1075 (2006).
[2]  J. Martín and C. Mijangos, Langmuir **25**, 1181 (2009).
[3]  W. S. Tung, R. J. Composto, R. A. Riggleman, and K. I. Winey, Macromolecules **48**, 2324 (2015).
[4]  S. Alexandris, P. Papadopoulos, G. Sakellariou, M. Steinhart, H. J. Butt, and G. Floudas, Macromolecules **49**, 7400 (2016).
[5]  H. Xiang, K. Shin, T. Kim, S. I. Moon, T. J. McCarthy, and T. P. Russell, Macromolecules **37**, 5660 (2004).
[6]  P. Huber, J. Phys. Condens. Matter **27**, 103102 (2015).
[7]  M. Rubinstein and R. H. Colby, *Polymer Physics* (2003).
[8]  N.-K. Lee, J. Farago, H. Meyer, J. P. Wittmer, J. Baschnagel, S. P. Obukhov, and A. Johner, EPL **93**, 48002 (2011).
[9]  Y. Yao, H.-J. Butt, G. Floudas, J. Zhou, and M. Doi, Macromol. Rapid Commun. **39**, 1800087 (2018).
[10] D. I. Dimitrov, A. Milchev, and K. Binder, Phys. Rev. Lett. **99**, 1 (2007).
[11] E. W. Washburn, Phys. Rev. **17**, 273 (1921).
[12] K. Kremer and G. S. Grest, J. Chem. Phys. **92**, 5057 (1990).
[13] R. Kumar, M. Goswami, B. G. Sumpter, V. N. Novikov, and A. P. Sokolov, Phys. Chem. Chem. Phys. **15**, 4604 (2013).
[14] A. Shavit and R. a Riggleman, Soft Matter **11**, 8285 (2015).
[15] R. Auhl, R. Everaers, G. S. Grest, K. Kremer, and S. J. Plimpton, J. Chem. Phys. **119**, 12718 (2003).
[16] N. Fries and M. Dreyer, J. Colloid Interface Sci. **327**, 125 (2008).
[17] J. Rafiee, X. Mi, H. Gullapalli, A. V.

Thomas, F. Yavari, Y. Shi, P. M. Ajayan, and N. a. Koratkar, Nat. Mater. **11**, 217 (2012).
[18] J. Alejandre, D. J. Tildesley, and G. A. Chapela, J. Chem. Phys. **102**, 4574 (1995).
[19] D. G. Legrand and G. L. Gaines, J. Colloid Interface Sci. **42**, 181 (1973).
[20] D. T. Wu, G. H. Fredrickson, J. -P Carton, A. Ajdari, and L. Leibler, J. Polym. Sci. Part B Polym. Phys. **33**, 2373 (1995).
[21] M. W. Matsen and P. Mahmoudi, Eur. Phys. J. E **37**, 78 (2014).
[22] R. J. Lang, W. L. Merling, and D. S. Simmons, ACS Macro Lett. **3**, 758 (2014).
[23] J. N. Israelachvili, *Intermolecular and Surface Forces: Third Edition* (2011).
[24] Y.-R. Huang, Y. Jiang, J. L. Hor, R. Gupta, L. Zhang, K. J. Stebe, G. Feng, K. T. Turner, and D. Lee, Nanoscale **7**, 798 (2015).